# An Ultra-compact Nanophotonic Optical Modulator using Multi-State Topological Optimization

Apratim Majumder[1], Bing Shen[1,i], Randy Polson[2], Trisha Andrew,[3] and Rajesh Menon[1]*

[1]Department of Electrical and Computer Engineering, University of Utah, Salt Lake City, Utah 84112, USA.

[2]Utah Nanofabrication Facility, University of Utah, Salt Lake City, Utah 84112, USA.

[3] Department of Chemistry, University of Massachusetts Amherst, Amherst MA 01003, USA.

*e-mail: rmenon@eng.utah.edu

**Abstract:** Optical modulators are one of the most important elements of photonic circuits. Here, we designed, fabricated and characterized a nanophotonic all optical modulator (NOM) in silicon that is an order of magnitude smaller than any previous silicon modulators, and exhibits an extinction ratio and operable spectrum that are an order of magnitude larger than alternatives. Simulations indicate that our device can provide about 9.5dB extinction ratio with a bandwidth of 10nm, centered at 1.55μm, corresponding to extinction ratio per device length of about 3.2dB/μm. Our device is fully compatible with standard complementary metal-oxide-semiconductor (CMOS) fabrication processes. The device is an example of a photonic device designed using multi-state topological optimization and will lead to a new class of ultra-compact and multi-functional active integrated-silicon devices.

**Introduction**

Optical communication systems typically incorporate photonic-integrated circuit (PIC) elements fabricated on the silicon-on-insulator (SOI) platform due to a number of advantages such as low power consumption, high efficiency, small footprint and compatibility with complementary metal-oxide-semiconductor (CMOS) fabrication processes [1-2]. An optical modulator is one of the most critical components in such a system [3]. Most commonly, optical modulation in silicon has been realized in the past using the plasma dispersion effect, where a change in the real and imaginary parts of the refractive index of silicon is caused by a change in its concentration of free charges [4]. However, this effect is fairly weak allowing for refractive index modulation of only 0.00332/V [5,6]. In general, modulators are characterized by the extinction ratio per modulation length, defined as $10\log(I_{max}/I_{min})/L$ dB/μm, where $I_{max}$ and $I_{min}$ are the intensities transmitted in the ON and OFF states, respectively, and L is the length of the device. Other important characteristics include the operating bandwidth, insertion loss and switching speed. Monolithic modulators, using direct carrier injection over p-i-n junction type devices were able to achieve 80% modulation over a 10μm device length [4], equivalent to 0.69 dB/μm [7]. Liu et. al. demonstrated modulation using carrier accumulation on either side of a dielectric layer inside a micrometer-sized waveguide. This work was improved by Liao et al. and a device capable of 20dB extinction ratio over a 3.45mm device length, corresponding to 5.8 X $10^{-6}$dB/μm was demonstrated. Although the device footprint was large, modulation speeds >1GHz were reported [8-10]. A smaller device using the same working principle with a 9dB extinction ratio achieved over 480μm device length, corresponding to 18.8 X $10^{-3}$dB/μm was later reported [10]. Modulation achieved using carrier depletion has also been reported [11-16], among which, Liao et. al. [12] reported the fastest data transmission at

---

[i] Current address: Macom Inc, New York.

40 Gbit/s with 1dB extinction ratio. However, since these devices typically employ Mach-Zehnder Interferometer (MZI) type phase shifters in their design, the device footprint is usually in the 1mm to 3.5 mm range, which corresponds to a range of 1 X $10^{-6}$dB/µm to 28.6 X $10^{-8}$dB/µm. Xu et. al. introduced active high-speed ring resonators [17-20] with extinction ratios as high as 15dB over device lengths of 12-15µm equivalent to about 1.25dB/µm. An important drawback of this device is a very narrow (<1 nm) optical bandwidth. An electro-optic polymer based modulator with 3dB modulation depth was recently demonstrated on a footprint of hundreds of microns, corresponding to < 3 X $10^{-2}$dB/µm [21]. Recently, there have been advances in modulators utilizing two-dimensional materials such as graphene [22-24], transition metal dichalcogenides [25-27], black phosphorus, etc. Of these, graphene performs exceptionally well, due to its high carrier mobility achieving extinction ratio per active length as high as 3dB/µm [24]. Different types of all-optical modulators have been demonstrated, such as saturable absorbers [28], optical limiters [29], polarization controllers [30] with large extinction ratios upto 27dB. However, these are discrete millimeter-scale devices and their on-chip counterparts have not been realized yet. A single atom plasmonic switch using silver with a 9.2dB extinction ratio was recently demonstrated [31]. However, extreme fabrication constraints as well as the use of non-CMOS compatible metals like silver and platinum required to realize this device limits its application. Most recently, Zhao et. al. demonstrated high all-optical modulation (60dB) using low power control signal, but on a millimeter-scale device, corresponding to < 0.06dB/µm [32].

In contrast to previous work, here we designed, fabricated and characterized an all-optical modulator that uses silicon and a CMOS-compatible photochromic molecule that achieves modulation of 3.2dB/µm in an area of only 3µm × 3µm with an operating bandwidth of about 10nm. The device may be fabricated in a single lithography step and can be fully CMOS compatible. The mechanism for optical modulation is based upon the change in refractive index of a photochromic molecule upon exposure to two different wavelengths of light [33-37]. Photochromes are low-density, soft molecular materials with widely adaptable processing requirements and can be introduced into a device stack at any point of a process flow. Further, films of the diarylethene photochromes used herein can be created using solvent-free physical vapor deposition, rendering this material CMOS compatible. A large extinction ratio per unit length is achieved due to the complex superposition of resonant guided modes that occur within our designed device as described below. Computationally designed nanophotonic devices have previously enabled a number of passive photonic functionalities [38-43] including free-space-to-waveguide couplers [38], optical diodes [40], polarization beamsplitters [41], densely packed waveguides [42], polarization rotators [43], etc. Here, we extend this computational method to active (multi-state) photonic devices using the example of an optical modulator.

**Design**

The device region (3µm × 3µm) is first divided into 30 × 30 square pixels, each of size 100nm × 100nm. Each pixel may be comprised of either silicon or of the photochromic material. Light of polarization state TE (electric field polarized in the plane of the device) is coupled into and out of the device via single-mode waveguides (see Fig. 1a). The photochromic material can be switched from a closed-ring state (n=1.62 at λ=1.55µm) to an open-ring state (n=1.59) and vice-versa by illuminating it to visible and UV wavelengths, respectively. This allows us to exploit a change in refractive index of 0.03 for the photochromic (multi-state) pixels, which is at least an order of magnitude higher than what is achievable in silicon using the plasma dispersion effect [3]. We first arbitrarily choose the closed-ring isomeric state of the photochromic molecule as the ON state and the open-ring isomeric state as the OFF state of the modulator. Details of the photochromic mechanism are further described in section 1 of the supplementary information [47]. A nonlinear optimization is used to design the topological distribution of the silicon/photochromic pixels with the goal of maximizing a figure of merit (FOM). Here, the FOM is defined as either the extinction ratio or the difference in transmission between the ON and OFF states of the modulator. We used a modified version of the direct-binary-search algorithm to implement the optimization (see section 2 of supplementary information) [47].

The optimized topology or pixel-distribution is shown in Fig. 1a. When the modulator is illuminated with UV light ($\lambda$~300nm), the photochromic pixels switch from the open-ring to the closed-ring state, and the modulator enables transmission of the light from the input to the output waveguides as illustrated by the simulated steady-state intensity distribution in Fig. 1b. On the other hand, when the modulator is illuminated with visible light ($\lambda$~633nm), the photochromic pixels undergo the opposite transformation, the modified refractive-index distribution results in low transmission of light from the input to the output (see simulated steady-state light distribution in Fig. 1c). It is well known that photochromic molecules of the diarylethene family, that we utilize here, can switch states in the time scale of picoseconds [44]. Hence, although not demonstrated here, this type of modulator may be used for fast optical switches. In each state, coupled resonant modes are excited by the incident field, which either results in relatively high or low transmission to the output waveguide depending upon the refractive-index distribution (see supplementary movie 1) [47]. We plotted the simulated relative transmission-efficiency spectrum, which is computed as the ratio of power in the output waveguide to that in a reference waveguide (without the modulator), in Fig. 1d. The extinction ratio of the modulator is defined as the ratio of the output power in the ON state to that in the OFF state for the same input power, and is plotted in Fig. 1e as a function of wavelength. The simulated extinction ratio is about 9dB at $\lambda$=1.55$\mu$m and a peak extinction ratio of 9.5dB is seen at $\lambda$=1.552$\mu$m. If we define the operating spectral bandwidth as the wavelength range over which the extinction ratio is maintained above 20% of its peak value, then we estimate the simulated bandwidth as ~10nm. This range is at least an order of magnitude larger than that achieved by devices that are similar in size [3]. In fact, only one device shows a larger spectral range (~20nm), but its area is about 3 orders of magnitude larger than our modulator [10]. Additional designs that were obtained with different figures-of-merit are further described in section 2 of the supplementary information [47].

Unlike traditional metamaterials, where control and modulation is achieved via symmetry breaking of periodic structures, our devices use free-form metamaterials, where the distribution of the silicon and photochromic pixels is determined using our computer algorithm. From the simulations, it is revealed that for the OFF state of the device, most of the light is scattered out of plane and onto the sides of the device, while decreasing the amount of light that is coupled into the output waveguide. For the ON state of the device, guided coupled resonant modes are generated that operate in unison to couple light into the output waveguide. As shown in Fig. 1d, the device behaves similar to a phase shifter, but over an extremely small active length. It is noted that our design technique results in a device that works using an ensemble effect, with a particular arrangement of pixels. Indeed, more than one solution (distribution of pixels) is possible with the device open to variable tuning in order to mitigate insertion losses or maximize different parameters such as transmission efficiency, extinction ratio, etc.

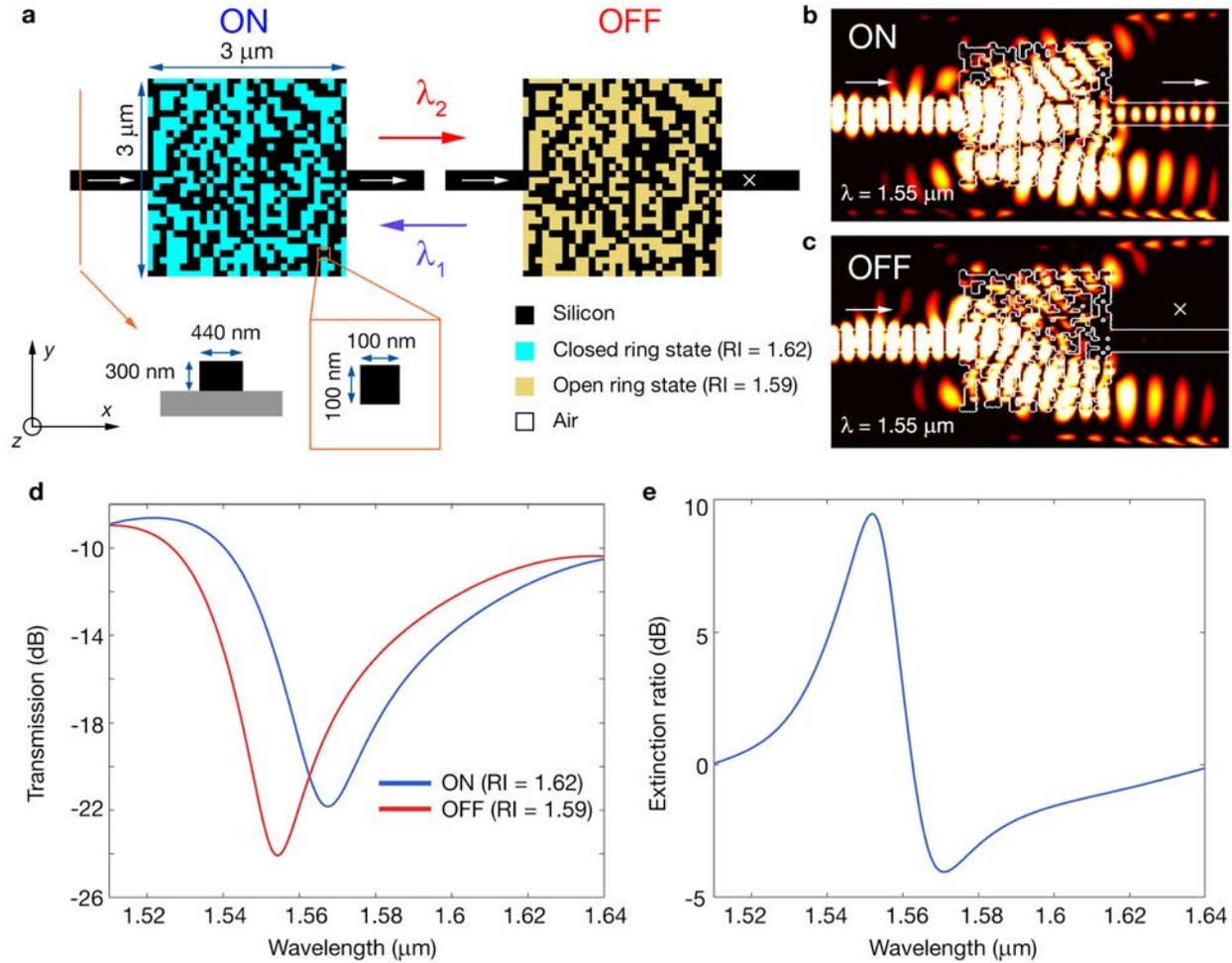

**Figure 1. The nanophotonic all-optical modulator (NOM). (a)** Schematic showing the geometry of the NOM device. Simulated steady-state intensity distributions for ON **(b)** and OFF **(c)** states of the device at the design wavelength of 1.55μm (see Supplementary Movie 1) [47]. The white arrows in **(a-c)** show the direction of propagation of light through the device. **(d)** Simulated relative transmission of the NOM as a function of wavelength. **(e)** Simulated extinction ratio as a function of wavelength, showing a peak of 9.5dB at λ=1.552μm and about 9dB at λ = 1.550μm, corresponding to a peak of 3.2dB/μm.

**Experiments**

The NOM was fabricated in an SOI wafer with silicon thickness of 300nm. The SOI wafer was prepared as described in section 2 of the supplementary information [47]. Due to our process constraints, we used poly-silicon for the device layer, resulting in higher than normal scattering losses [45]. As mentioned earlier, the device may be fabricated in a single lithography step, since the height of the pixels and that of the waveguides is the same as that of the top silicon layer. However, due to our limited access to high-resolution lithography tools, we opted for a two-step patterning process. All coarse features were first patterned using maskless lithography along with alignment marks (see section 3 of the supplementary information [47]). Then, the fine features (layout shown in Fig. 2a) were patterned using focused ion-beam lithography. The photochromic material was spin coated directly onto the device after all lithography was complete. On-chip polarizers were included in our devices to ensure that the input polarization state was TE. Reference devices (waveguides with no modulator) were also included on the same wafer for reference measurements. Scanning-electron micrographs of the patterned device before and after coating with the photochromic material are shown in Figs. 2b and 2c, respectively. Finally, the fabricated devices were characterized using a setup that we have described previously [30-34] (also included in section 4 of the supplementary information [47]). Switching between the ON and OFF states was achieved by illuminating the entire chip with visible and UV light from above. As mentioned earlier, the transmission efficiency was computed relative to the reference device to account for any inadvertent polarization rotation that occurs during the free-space-to-waveguide coupling and during the taper from multi-mode to single-mode waveguides.

**Results**

We plotted the transmitted power as a function of wavelength in Fig. 2d for the fabricated NOM. Fig. 2e shows the extinction ratio as a function of wavelength. Although the measured spectra are somewhat different than simulations, modulation is observed between the two states. The peak measured extinction ratio was about 6dB at $\lambda=1.55\mu m$, equivalent to 2db/$\mu m$. The measured operating bandwidth is ~40nm.

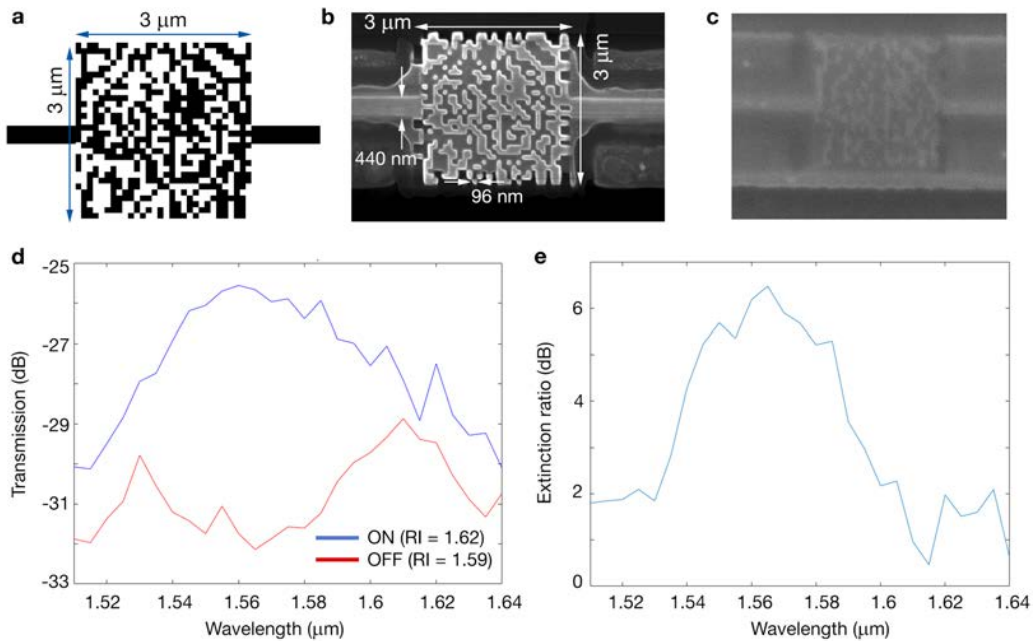

**Figure 2. Experimental results. (a)** Layout of the NOM geometry. White regions represent etched silicon. Scanning-electron micrographs of the fabricated device **(b)** before and **(c)** after over-coating with the photochromic material. Measured relative transmission and extinction ratios as functions of wavelength are plotted in **(d)** and **(e)**, respectively.

In order to understand the difference between experiments and simulations, we performed a numerical analysis of the impact of various fabrication errors on the performance of the device. The results are summarized in Fig. 3. The 3 main fabrication errors studied were: the silicon thickness (Fig. 3a), the size of the pixel (Fig. 3b) and misalignment between the input/output waveguides and the modulator (Fig. 3c). From this analysis, we can surmise that a drop in extinction ratio from 9dB to 6dB at $\lambda=1.55\mu m$ is possible due to either a 10 nm difference in silicon thickness, a 5 nm change in pixel size or a misalignment error of about 100nm, or some combination of these errors. In addition, the use of polysilicon for the device layer increases scattering losses (see section 3 of the supplementary information) [47]. Similar analyses of other modulator designs are also included in section 5 of the supplementary information [47].

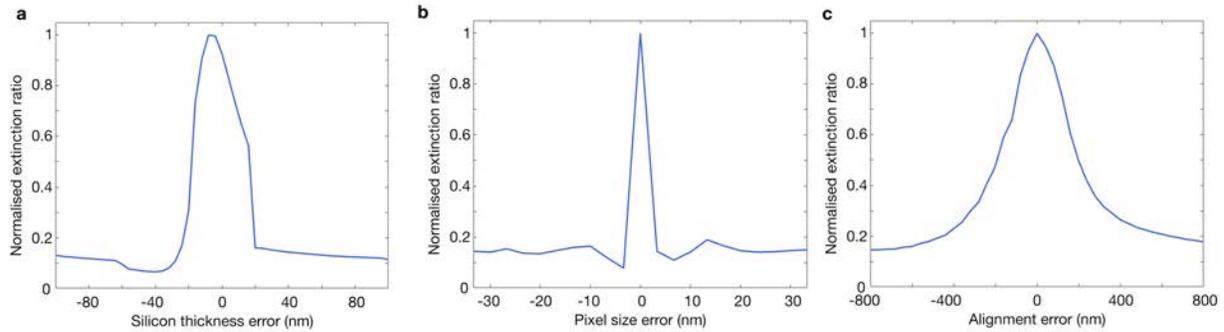

**Figure 3. Effect of fabrication errors.** Simulated extinction ratio at λ=1.55μm as a function of errors in **(a)** the silicon layer thickness, **(b)** pixel dimensions and **(c)** alignment between the input/output waveguides and the modulator.

The refractive index of the photochrome can be changed continuously from 1.62 to 1.59 (λ=1.55μm), and back by exposing to the appropriate wavelengths of illumination. This allows us to continuously change the transmission of the NOM. The simulated (normalized) transmission as a function of refractive index of the photochrome pixels is shown in Fig. 4a. In order to experimentally characterize this, we first spin-coated a thin uniform layer of the photochrome on a silicon wafer. Then, this layer was exposed to visible light (λ~633nm) until the photostationary state is reached. The refractive index of this layer was measured and recorded using a spectroscopic ellipsometer (Woollam). Then, the sample was exposed to a UV light source (λ~300nm) and the refractive index was monitored as a function of time, resulting in the plot of refractive index vs exposure dose (Fig. 4b). Next, the transmission of the NOM was also measured as a function of the exposure dose in the same manner (repeated 3 times) and the averaged results are plotted in Fig. 4c. By combining Figs. 4b and 4c, we can obtain the transmission of the NOM as a function of the refractive index of the photochrome pixels and is shown in Fig. 4d. It is noted that the measured curve closely matches the simulated one. Therefore, our NOM is also able to function as an analog modulator.

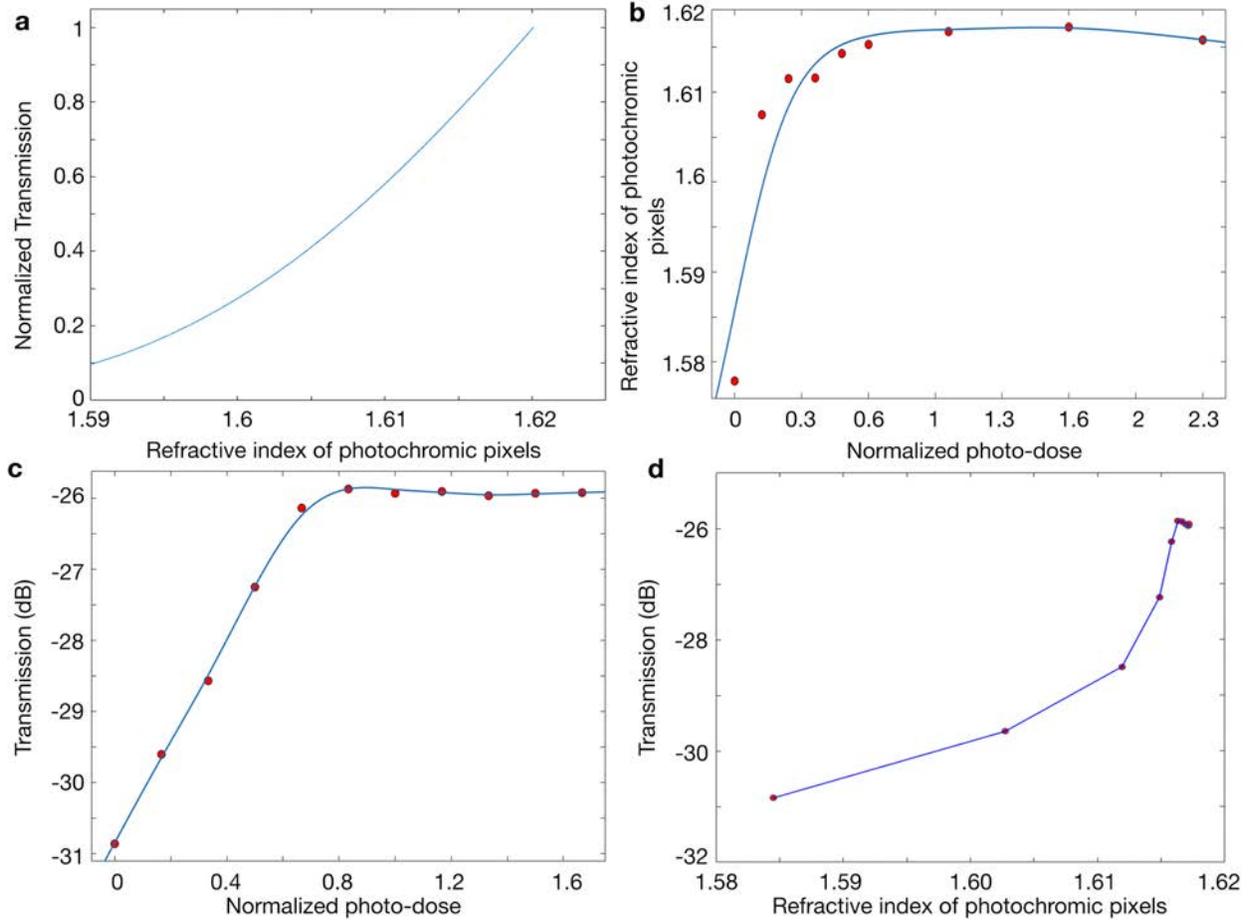

**Figure 4. Analog modulation. (a)** Simulated NOM transmission as a function of the refractive index of the photochromic pixels. **(b)** Measured refractive index of a uniform photochromic film and **(c)** measured NOM transmission as a function of photo-dose (UV, λ~300nm). **(d)** Measured NOM transmission as a function of the refractive index of the photochromic pixels.

**Conclusion**

In summary, we designed, fabricated and characterized a nanophotonic all optical modulator using the concept of computationally designed digital metamaterials and photochromes, that is at least an order of magnitude smaller in area than the smallest equivalent device demonstrated before. Our device also provides an order of magnitude increase in the operable bandwidth and is to our knowledge, the first demonstration of a metamaterial-based integrated optical modulator. We also present the application of photochromes to this field that offer a refractive index modulation ten times greater than that found in silicon. Our simulations indicate that an extinction ratio per length of about 3.2dB and an operating spectrum of 10nm is possible. We were able to experimentally show an extinction ratio of 6dB. Furthermore, our device can be CMOS compatible for large-scale integration. It is important to note that modulation bandwidth is constrained by the fundamental switching speed of the photochrome, i.e. the reaction rates of the photochemical transformations between the open and closed forms. For the diarylethene photochrome used here, the slowest reaction is known to be the transformation of the closed to the open form. The reaction rate of this transition has been measured to be on the order of 10^10 transformations per second [46], implying that 10 GHz modulation may be achieved in principle.

**Acknowledgements**

The authors thank T. Olsen, S. Pritchett, B. Baker, N. Mohammad and A. Banerjee for assistance and advice with the fabrication. This work made use of the University of Utah shared facilities of the Micron Technology Foundation Inc. Microscopy Suite sponsored by the College of Engineering, Health Sciences Center, Office of the Vice President for Research and the Utah Science Technology and Research (USTAR) initiative of the State of Utah. This work made use of University of Utah USTAR shared facilities, supported in part by the MRSEC Program of the National Science Foundation (NSF, award no. DMR-1121252).

**Author contributions**

A.M., B.S. and R.M. conceived and designed the experiments. B.S. and A.M. performed the simulations. A.M. and R.P. fabricated the device, contributed materials/analysis tools and performed characterizations. A.M., B.S. and R.M. analysed the data. A.M. and R.M. wrote the paper.

**Additional information**

Supplementary information is available in the online version of the paper. Correspondence and requests for materials should be addressed to R.M.

**Competing financial interests**

The authors do not disclose any competing financial interests.

Supplementary Information

# An Ultra-compact Nanophotonic Optical Modulator using multi-state topological optimization

Apratim Majumder[1], Bing Shen[1,i], Randy Polson[2] Trisha Andrew,[3] and Rajesh Menon[1]*

[1]Department of Electrical and Computer Engineering, University of Utah, Salt Lake City, Utah 84112, USA. [2]Utah
[2]Nanofabrication Facility, University of Utah, Salt Lake City, Utah 84112, USA.

[3] Deparment of Chemistry, University of Massachusetss Amherst, Amherst MA XXXX, USA.

*e-mail: rmenon@eng.utah.edu

1. **Details of the photochromic material**

We used a specific photochromic material as the switchable pixels of the nanophotonic all-optical modulator (NOM). The photochrome used in this work belongs to the diarylethene family. Specifically, it is the 1,2-bis(5,5' –dimethyl-2,2'-bithiophen-4-yl) perfluorocyclopent-1-ene molecule (otherwise referred to as BTE). We have previously used this material in demonstrating super-resolution photolithography [1-4]. The chemical structure of the molecule is shown in Fig. S1a. The molecule has the ability to reversibly transition between two isomeric states with different molecular structures, based on the wavelength of the photon absorbed by each state. We call these two isomers, the "open" and the "closed" forms, based on the structure of the central benzene ring in the respective isomer. The open form converts to the closed form by absorbing photons around 320 nm wavelength and the reverse state transition is initiated by the closed form absorbing photons at 615 nm. We use out-of-plane illuminations at these two wavelengths to induce the state changes of the photochromic pixels. The UV-Vis spectrophotometric analysis of the molecule showing its absorbance spectra is presented in Fig. S1b. At the same time, these two isomeric forms possess different refractive indices (real part of refractive index, $n_{Open}$ for the open form and $n_{Closed}$ for the closed form) at the communication wavelength $\lambda = 1.55 \mu m$. This has been reported previously [5] and we have independently confirmed this observation by measuring the refractive indices of the two forms using ellipsometric analysis on a Woollam Variable Angle Spectroscopic Ellipsometer (VASE). This is shown in Fig. S1c. For the NOM, the two different values of the real part of the refractive index of the photochrome at $\lambda = 1.55 \mu m$ govern the two states ("ON" and "OFF") of the modulator. The difference of the two values of refractive index at $\lambda = 1.55 \mu m$ ($\Delta n = n_{Closed} - n_{Open} = 1.62 - 1.59$) is 0.0301 as shown in Fig. S1d. This is significantly larger than the refractive index difference available in silicon by electrical effects by at least an order of magnitude [6]. Additionally, the photochromic material has no absorbance in the range $\lambda = 1.5 \mu m$ to 1.6 µm as seen in the ellipsometry data presented in Fig. S1e, which indicates that the NOM does not function via any absorbance of the input signal. The photochromic pixels in the NOM are defined by first using FiB milling to create empty pixels by etching away and thereby removing silicon, followed by spin coating the photochrome to fill these empty pixels and thereby create the photochromic pixels. In order to spin coat this material, the photochromic molecules are first dissolved by 93.63 weight percent in a polymer matrix of a 30 mg/mL solution of polystyrene dissolved in toluene. The process of synthesizing the molecules have been described elsewhere [1].

---

[i] Current address: Macom Inc, New York.

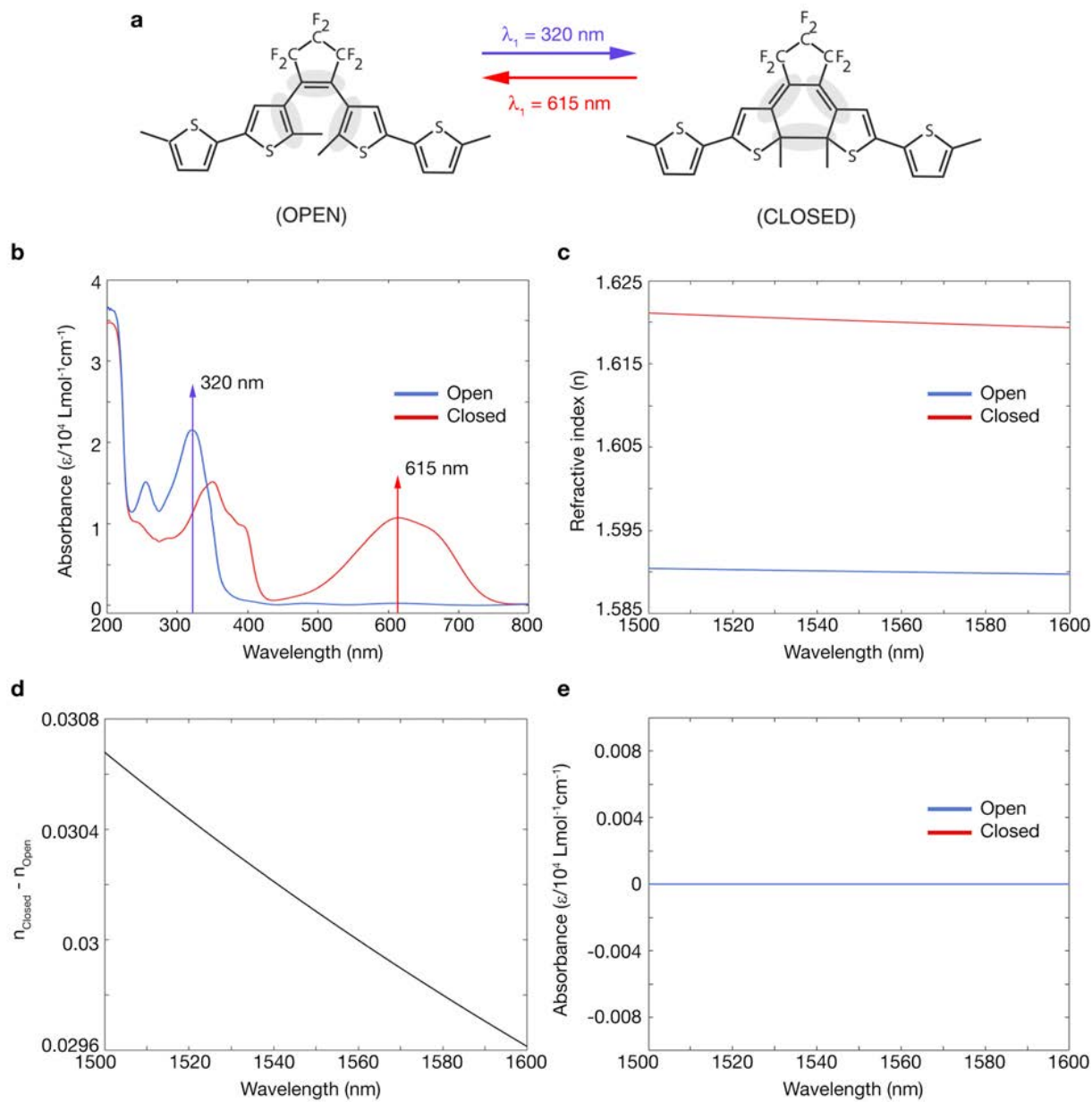

**Figure S1. Photochromic material. (a)** Molecular structures of the open and closed isomers of the photochromic molecule showing the state change due to absorbance of UV and visible photons [1] **(b)** UV-Vis spectrophotometry data showing the absorbance of the photochromic material with peaks at 320 nm for the open form and 615 nm for the closed form **(c)** Ellipsometry data showing two different refractive indices for the two isomers of the molecule. Independantly measured and confirmed with [5] **(d)** Difference in the refractive index of the two forms **(e)** UV-Vis spectrophotometry data showing no absorbance of the photochromic material in the wavelength range 1500 to 1600 nm.

## 2. Design

Similar to our previous work [7-12], here, we used a modified version of the direct-binary-search (DBS) algorithm [13,14] to design the nanophotonic all optical modulator. One of the main advantages of using this algorithm over other similar yet distinct optimization techniques like generic algorithm and inverse design [15,16] is the fact that DBS has the ability to discretize the space over which the desired electromagnetic phenomenon takes place and thereby produce structures with discrete pixels, thereby facilitating fabrication in the nanoscale using conventional techniques like focused ion beam (FiB) milling, as we used here, or electron beam (e-beam) lithography, which would be a more idea choice of tool. Discrete pixels make our devices easier to fabricate, compared to continuous structures as produced by the result of inverse design algorithms [15]. In DBS, a given region of space is discretized in to a number of pixels. In this case, the region of space was a 3μm × 3μm area, discretized into 30 × 30 square pixels. Each pixel thus has dimension of 100nm × 100nm which also defines the size of the smallest feature that needs to be fabricated for our device. This is well within the reach of prevalent nanofabrication techniques like the FiB and e-beam lithography. We also chose to fix the maximum height of the pixels at 300 nm in order to facilitate the FiB milling and keep the etched sidewalls as straight as possible. We have noticed that beyond 300 nm, it is extremely difficult to produce straight sidewalls. Following this, the DBS algorithm performs a nonlinear optimization to arrive at a specific pattern for the pixels that achieves the desired electromagnetic phenomenon, which in this case is intensity modulation of an input signal. One of the disadvantages of DBS is its high computation time. Hence, we chose to parallelize the process using cloud computing on Amazon Web Services servers. The other drawback is that at times DBS tends to converge prematurely due to its sensitivity to the initial conditions. As a result, we repeated the optimization process with a number of random initial designs generated by Matlab. The design presented in this manuscript has the largest figure of merit (FOM) among a number of designs with different starting points.

For the NOM, there are two types of pixels. These are silicon and the photochromic material. The latter again may exist in one of its two states with different refractive indices as mentioned in the previous section. We define a figure of merit (FOM) as the average transmission efficiency of the input signal through the NOM device. A random starting pattern for the pixel layout was generated by Matlab and then passed into the DBS algorithm. The FOM was calculated for this pattern with the photochromic material in its two states, open and closed, as mentioned in the previous section, with two different values of refractive index in the two states, 1.59 and 1.62, respectively. The FOM when the photochromic material is in the open state is '$FOM_{Open}$' and when it is in the closed state is '$FOM_{Closed}$'. Without any loss in generality, we attributed the high transmission state of the switch to when the photochromic pixels are in the closed state. Hence, a final FOM is calculated as $FOM_{Final} = FOM_{Closed} - FOM_{Open}$. Next, the pixel is toggled to the other material state, i.e. if the pixel was silicon in the previous step, it is toggled to photochromic material and if it was the latter, then it is toggled to silicon. Again, $FOM_{Final}$ is calculated. If the new value of $FOM_{Final}$ is larger than the previous value, then the current pixel material state is retained, otherwise switched back. Then the next pixel is toggled and this iterative optimization of the process continues. One iteration ends when all 900 pixels have been toggled. The total optimization terminates when the FOM does not improve beyond a predefined threshold (<0.5% in this case). We used finite-difference-time-domain (FDTD) method for simulating the electromagnetic fields in the devices and calculating the FOM, similar our previous work [7-11].

We designed three devices with slightly varying definitions of $FOM_{Final}$. The schematic of the geometry of the three designs with their respective simulated transmission and extinction ratios are shown in Fig. S2. A comparative study of the transmission and extinction ratios of all three designs is shown in Fig. S3. For the first device, the final FOM was defined to simply maximize the extinction ratio. For the second device, it was defined to maximize the transmission at λ = 1.55µm. Hence, it can be seen that the peak of the extinction ratio (7.7dB) of the first device shifts from 1.544µm to 1.550µm in the second device. For the first device, the extinction ratio at 1.550µm was 6dB. Finally, for the third device, both these parameters were maximized to obtain an extinction ratio of 9dB at 1.550µm and a peak of 9.465dB at 1.552µm. This shows that our device design is completely computationally driven and possesses enough generality that the design process may be easily adapted to design different types of optical modulators in the future, with specific device performance tuned to the nanometer scale.

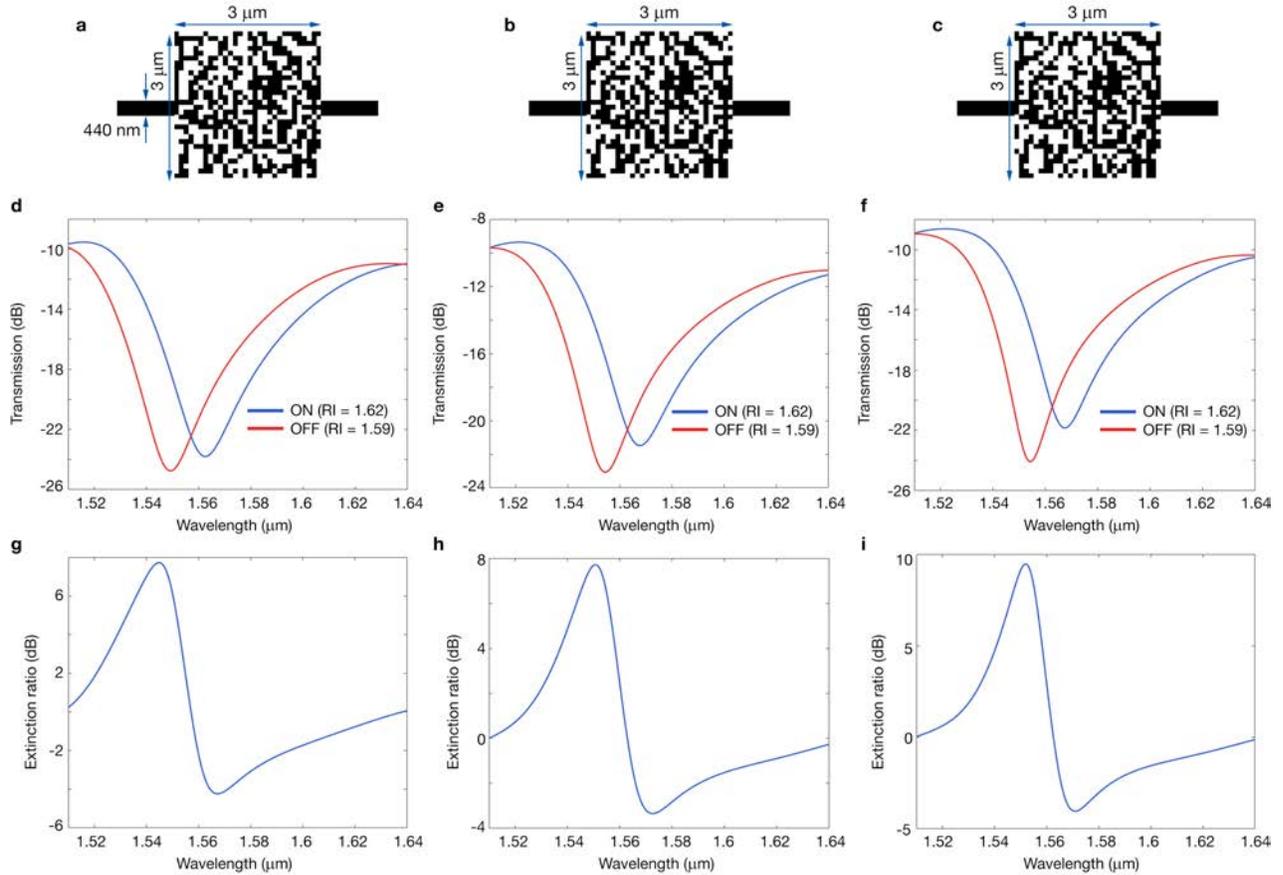

**Figure S2. Nanophotonic all optical modulator (NOM) designs.** Three designs for NOM devices. **(a-c)** Geometries of three devices. Black pixels are silicon and white pixels indicate where photochromic material is to exist in either of its two states. **(d-f)** Simulated transmission as a function of wavelength. **(g-i)** Simulated extinction ratio as a function of wavelength. Transmission and extinction ratio for device shown in **(a)** are presented in **(d)** and **(g)** respectively, those for the device shown in **(b)** are presented in **(e)** and **(h)** respectively and those for the device shown in **(c)** are presented in **(f)** and **(i)** respectively.

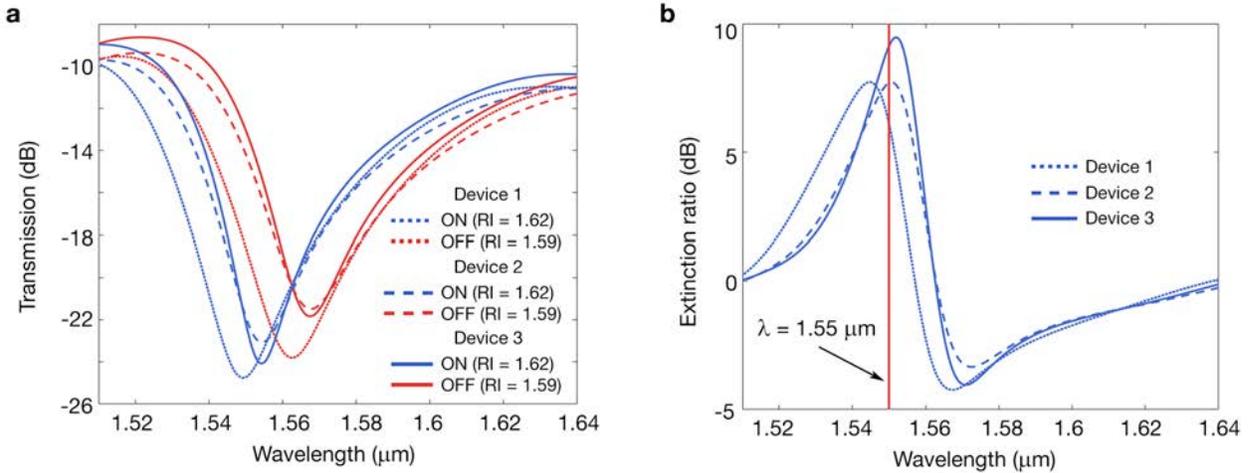

**Figure S3. Comparison of designs for NOM device.** Comparison of the **(a)** transmission and **(b)** extinction ratio as a function of wavelength for three designs for NOM devices. The vertical red line in **(b)** indicates λ = 1.55µm.

### 3. Detailed sample preparation method

We fabricated the NOM device on the conventional platform of a silicon-on-insulator (SOI) wafer. As per our design constraint, the thickness of the top silicon layer was required to be 300 nm. However, since we were unable to procure SOI wafers with this exact silicon layer thickness at the time, we resorted to preparing our own wafer. In order to do this, first a 4-inch silicon wafer (University Wafers) was cleaned using acetone, methanol and isopropanol. Subsequently, the wafer was etched using oxygen plasma on a Technics PE-II to remove all surface contaminants. Next, we deposited a 3 µm thick silicon oxide ($SiO_2$) layer on the surface of this wafer using plasma enhanced chemical vapor deposition (PECVD) technique on an Oxford Plasmalab 80+. Lastly, polycrystalline silicon (PolySi) of thickness ~ 300 nm was deposited in an Expertech furnace using the method of low power chemical vapor deposition (LPCVD), deposited at 630°C at a deposition rate of ~ 11.5 nm/min. PolySi exhibits large granularity as shown in Fig. S4a. Unfortunately, this leads to undesired light scattering and contributes majorly to losses in the device. Since the height of the pixels in our device is the same as that of the input and output single mode waveguides, which is same as the height of the top silicon layer, our entire device with all waveguides and tapers may be fabricated in one single step. However, we chose a two-step fabrication process for defining the pixels of the device since we did not have access to a projection lithography or electron beam lithography tool that is capable of fabricating features sizes in the range of tens of millimeters to hundreds of nanometers. In the first step, we fabricated all the larger waveguides with tapers down to a feature size of 3 µm. For this, the SOI wafer was first cleaned using acetone, methanol, isopropanol and subsequent oxygen plasma etching. Then a thin layer of Hexamethyldisilazane (HMDS) was spun on top of the wafer as an adhesion promoter for the photoresist layer. The wafer was then coated with a ~ 500 nm thick layer of Shipley S1805 photoresist. Photolithography was carried out using the direct laser writing (DLW) optical lithography tool Heidelberg µPG 101 to pattern all multi-mode waveguides and tapers. The photoresist was then developed by immersing the sample in AZ 1:1 developer for 60 seconds, followed by rinsing in DI water for 60 seconds to clean the sample of residual developer. The patterned photoresist layer was then used as an etch mask to etch the top silicon layer and thereby transfer the pattern to it. For this purpose, we used deep reactive-ion-etching (DRIE) on an Oxford 100 etcher using a modified Bosch process in order to achieve straightest possible sidewalls and lowest roughness. We used a 3 second etch step of $SF_6$ gas flowing at 40 sccm and $C_4F_8$ gas flowing at 17.5 sccm, followed by an immediate 2 second passivation step of only $C_4F_8$ gas flowing at 17.5 sccm, while operating the chamber at $0^{\circ}C$. This gave us a silicon etch rate of ~ 10 nm/min. In the second step of the patterning process, dual-beam focused ion beam (FiB) milling on a FEI Helios NanoLab 650 was used to define the "empty" pixels of the NOM (~ 100 nm x 100 nm), where silicon would be removed, to be later filled with photochromic material. Thus, the device along with tapers and all single-mode waveguides (440 nm width) could be fabricated in this step, as shown in the scanning electron micrograph in Fig. S4f. The beam current used was 7.7 pA with a fluence of 800 $C/m^2$. Standard alignment marks and fiducials were used to align the patterns in the two lithography steps and to correct for systematic beam drift caused by charging effect, change of surrounding temperature and other undesirable factors. Reference devices for normalization including the same tapers but no device and on-chip polarizers (Fig. S4b-e) for setting the input polarization state were also fabricated by this same process on the same wafer. Next, the empty pixels defined up till this point needed to be filled with photochromic material. We resorted to spin coating the photochromic material onto the sample at this point, thereby filling the empty pixels. In order to do this, molecules of the BTE in the powder form were dissolved in a 30 mg/mL solution of polystyrene dissolved in toluene, by 93.63 weight percentage. This solution was prepared by stirring in an ultrasonicator for 4-5 hours. The sample with the waveguides and the NOM was first coated with a monolayer of HMDS to facilitate adhesion. Next, the photochromic solution was spin coated at 1000 rpm for 60 seconds with a prior 500 rpm spread for 5 seconds. We first confirmed that the photochromic material covered the entire device region using an optical microscope. We then proceeded to perform the measurements and lastly, obtained scanning electron micrographs of the device region showing the photochromic pixels filled with this material. This imaging was reserved for the very end because the scanning electron beam tends to degrade the photochromic material. Moreover, we needed to coat the sample with a thin 2-3 nm layer of AuPd to facilitate the SEM imaging. Both these steps tend to affect the photochromic material in adverse ways. The scanning electron micrograph as shown in Fig. S4g confirm that the photochromic pixels were properly covered and filled by the material. The image quality is poor since at 300nm, the layer is thick enough to hinder the SEM imaging process. In order to check if the spinning of the photochromic material breaks the fragile silicon pixels, we lastly immersed the devices into a toluene ultrasonic bath for 30 minutes to dissolve the photochromic layer and thereby remove it. We were able to obtain a scanning electron micrograph where the photochromic layer had peeled but was still sticking to the edge of the device as shown in Fig S4h. It also shows that the silicon pixels were intact and survived the photochromic layer spin coating process.

We resorted to design and fabricate three devices as detailed in the previous section.

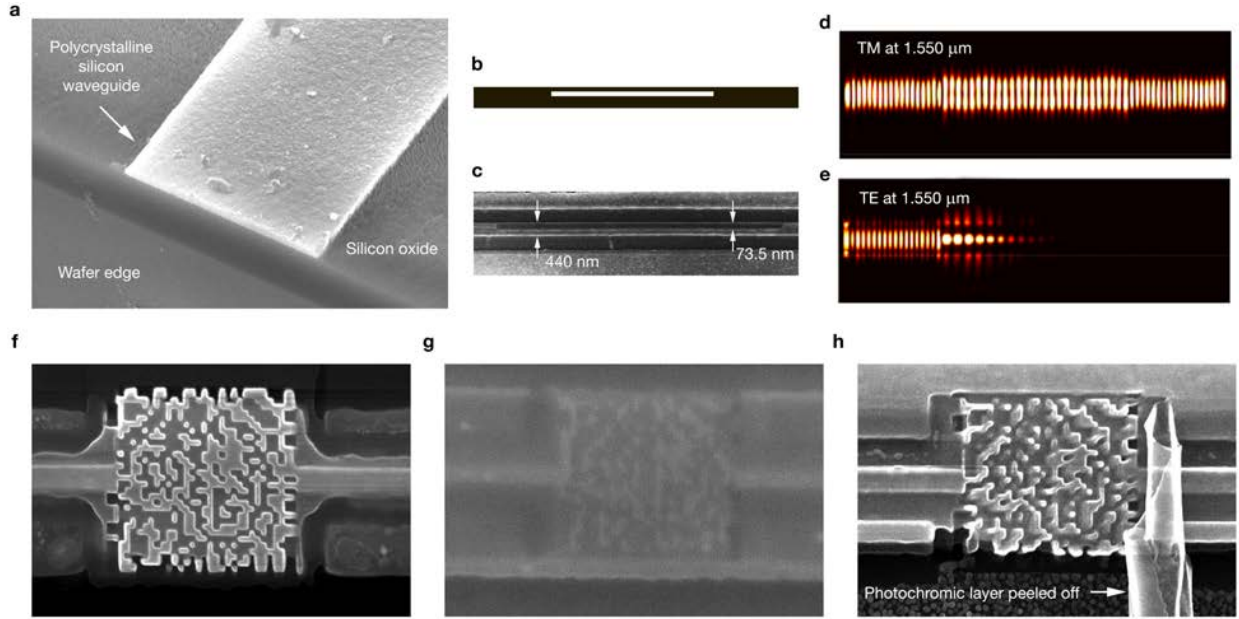

**Figure S4. Device fabrication. (a)** Scanning electron micrograph showing the granularity present in the polysilicon used as the top device layer of our self-prepared SOI wafer. **(b)** Schematic and **(c)** scanning electron micrograph of onchip polarizer used to set the input polarization state. In **(b)**, black is silicon and white is air. **(d, e)** Simulated steady-state light field distribution for the onchip polarizer showing TE polarized light being effectively blocked while TM polarized light being transmitted unperturbed. Scanning electron micrographs of the NOM showing the device after **(f)** the FiB milling step indicating the "empty" pixels, **(g)** the photochromic material spin-coating step and **(h)** the photochromic material removal step showing a thin strip of the material still stuck on to the edge of the device.

## 4. Characterization method

The setup used to characterize the fabricated NOM device is schematically shown in Fig. S5a. A standard single-mode lensed fiber was used at the input and a standard multi-mode fiber was used at the output to couple light into and from the multimode waveguides that then taper down to the single mode waveguides coupling in and out of the NOM device. The source was a Hewlett-Packard 8186F tunable laser source emitting in the range 1.508-1.640 µm. Similar to our previous work [7-11], we used an on-chip polarizer to set the polarization state of the input light. Fig. S4b,c show the geometry and the scanning electron micrograph of the on-chip polarizer used in the experiments. It has a simple design consisting of a straight single-mode waveguide with a horizontal air slot that is 70 nm offset from the center of the waveguide. As the simulated steady state light fields show in Fig. S4d and e, the onchip polarizer allows the transmission of the TM mode while effectively blocking the TE polarization. The detailed characterization process is as follows:

First, the single-mode input lensed fiber is aligned to the on-chip polarizer input. Butt-coupling is used to couple light from the lensed fiber to the waveguide that couples to the on-chip polarizer. The output is collected using a multi-mode lensed fiber. The fiber polarization controller PC1 is adjusted to set the input polarization state to TE by monitoring the power at the detector. The on-chip polarizer only allows the TM mode to pass through efficiently while blocking the TE mode. Hence, we could set the TE mode for the input by minimizing the power received at the detector.

Next, the input lensed fiber and output fiber could be moved to illuminate a straight reference waveguide with no device but all similar tapers or one with the NOM device in place. Once aligned, the measurements were recorded. For the NOM device, as shown in Fig. S5b, the device is first illuminated by a source with its emission centered at 320 nm to convert the photochromic pixels to the closed state, with refractive index 1.62 and also high transmission for the switch. The output is collected. Next the device is illuminated by a source centered at 615 nm to convert the photochromic pixels to the open state, with refractive index 1.59 and thus, low transmission for the switch. The output is again collected. Thus, we could measure all the NOM devices fabricated.

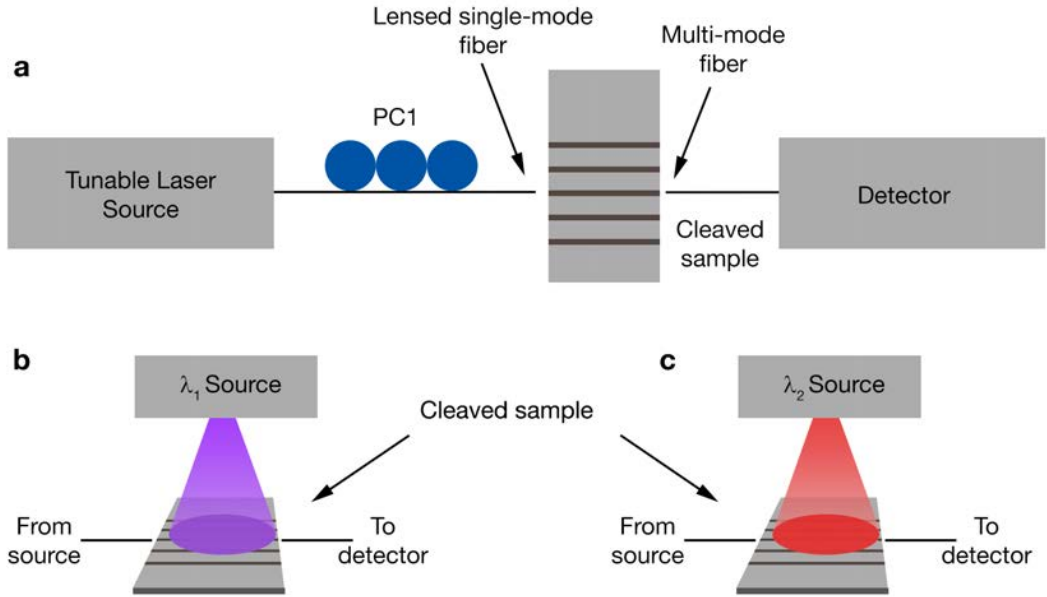

**Figure S5. Measurement setup. (a)** Schematic of the measurement setup used to characterize the NOM device. The input signal is provided via a standard single-mode lensed fiber and the output is collected using a standard multi-mode fiber, in both cases using butt-coupling. **(b)** Schematic of the switching procedure showing the sample being illuminated by two sources emitting photons that can induce state changes in the photochromic material. Measurements are collected after each switching step.

5. **Fabrication tolerance**

Fabrication errors are unfortunately unavoidable during our fabrication of the NOM device. There are primarily three sources of these errors: (a) variation in the silicon layer thickness, since we deposited the polySi layer ourselves while preparing the custom SOI wafer, (b) pixel dimension error introduced by non-uniform FiB pixel writing, and (c) misalignment between the input and output waveguides and the device, introduced by the two-step lithography process. In this section, we have provided detailed analyses of the effect of these errors on the performance of the NOM device. All the data is presented at $\lambda = 1.55 \mu m$.

  5.1. Variation in the silicon layer thickness
      The height of the pixels as defined by the Si layer thickness is uniformly varied for a range of 100 nm higher and lower than the targeted thickness of 300 nm and the effect of this on the performance of the device was simulated. The degradation in the extinction ratio can be clearly

observed in Fig. S7a, d and g, for the first, second and third device, respectively. We suspect this to be the major error plaguing our fabricated devices.

5.2. Pixel dimension error

The dimension of the NOM device was scaled and the degradation in the extinction ratio of the device was plotted as a function of this scaling factor as shown in Fig. S7b, e and h for the first, second and third device, respectively. The simulated extinction ratio falls off sharply from the peak that is achieved when the pixel dimension error is 0, i.e. no dimension error is present.

5.3. Alignment error

The lithography process that we used to fabricate the NOM device had two steps that needed to be precisely aligned with each other. Otherwise, the input and output waveguides would not be aligned to the device. This would introduce undesired mode evolution inside the device and degrade its performance. We simulated this effect for a misalignment range of 0-200 nm on the extinction ratio of the first, second and third device, as shown in Fig. S7c, f and I, respectively. However, this error may be readily solved using a single step fabrication method like using electron-beam lithography to fabricate all waveguides and device together.

6. **Animations description**

Two animations are provided as supplementary information showing the simulated time evolution of the electric field pattern within the nanophotonic all optical modulator for the on and the off states. The dimensions of the device are same as described in the manuscript. Light travels from left to right. The contour plot of the permittivity distribution of the device is shown using white lines.

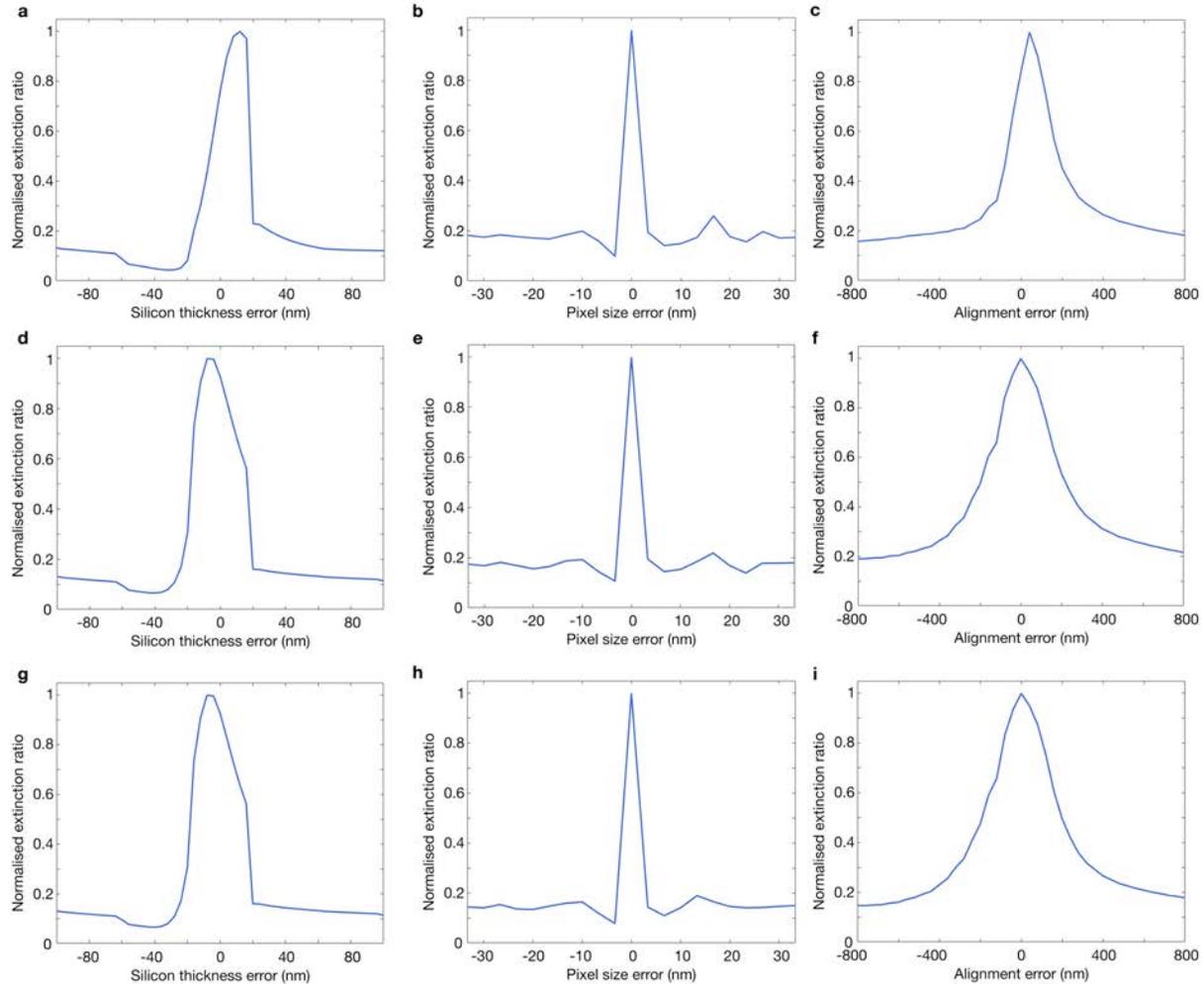

**Figure S7. Effect of fabrication errors.** Simulated extinction ratio at λ=1.55µm as a function of errors in the silicon layer thickness **(a, d and g)**, pixel error dimension **(b, e and h)** and misalignment between input and output waveguides and the NOM device. **(a-c)** are for the first device **(d-f)**, for the second device and **(g-i)** are for the third device.